\documentclass[%
 reprint,
 superscriptaddress,
 showpacs,preprintnumbers,
 amsmath,amssymb,
 aps,
 prb,
]{revtex4-1}

\usepackage{color}
\usepackage{graphicx}

\begin{document}

\title{Magnetic properties  of single-crystalline CeCuGa$_{3}$}

\author{Devang A. Joshi}
\affiliation{Institut f\"{u}r Festk\"{o}rperphysik, Karlsruhe Institute of Technology, D-76021 Karlsruhe, Germany}%
\author{P.~Burger}
\affiliation{Institut f\"{u}r Festk\"{o}rperphysik, Karlsruhe Institute of Technology, D-76021 Karlsruhe, Germany}%
\author{P.~Adelmann}
\affiliation{Institut f\"{u}r Festk\"{o}rperphysik, Karlsruhe Institute of Technology, D-76021 Karlsruhe, Germany}%
\author{D.~Ernst}
\affiliation{Institut f\"{u}r Festk\"{o}rperphysik, Karlsruhe Institute of Technology, D-76021 Karlsruhe, Germany}%
\author{T.~Wolf}
\affiliation{Institut f\"{u}r Festk\"{o}rperphysik, Karlsruhe Institute of Technology, D-76021 Karlsruhe, Germany}%
\author{K.~Sparta}
\affiliation{Institute of Crystallography, Rheinisch-Westf\"{a}Iische Technische Hochschule (RWTH) Aachen, J\"{a}gerstrasse 17/19, D-52056 Aachen, Germany}%
\author{G.~Roth}
\affiliation{Institute of Crystallography, Rheinisch-Westf\"{a}Iische Technische Hochschule (RWTH) Aachen, J\"{a}gerstrasse 17/19, D-52056 Aachen, Germany}%
\author{K.~Grube}
\affiliation{Institut f\"{u}r Festk\"{o}rperphysik, Karlsruhe Institute of Technology, D-76021 Karlsruhe, Germany}%
\author{C.~Meingast}
\affiliation{Institut f\"{u}r Festk\"{o}rperphysik, Karlsruhe Institute of Technology, D-76021 Karlsruhe, Germany}%
\author{H. v. L\"{o}hneysen}
\affiliation{Institut f\"{u}r Festk\"{o}rperphysik, Karlsruhe Institute of Technology, D-76021 Karlsruhe, Germany}%
\affiliation{Physikalisches Institut, Karlsruhe Institute of Technology, D-76031 Karlsruhe, Germany}%
\date{\today}

\begin{abstract}
The magnetic behavior of single-crystalline CeCuGa$_{3}$ has been investigated. The compound forms in a tetragonal BaAl$_4$-type structure consisting
of rare-earth planes separated by Cu-Ga layers. If the Cu-Ga site disorder is reduced, CeCuGa$_{3}$ adopts the related, likewise tetragonal BaNiSn$_3$-type structure, in which the Ce$^{3+}$ are surrounded by different Cu and Ga layers and the inversion symmetry is lost. In the literature conflicting reports about the magnetic order of CeCuGa$_{3}$ have been published. 
Single crystals with the centrosymmetric structure variant exhibit ferromagnetic order below $\approx $4\,K with a strong planar anisotropy. The magnetic behavior above the transition temperature can be well understood by the crystal-field splitting of the 4$f$ Hund's rule ground-state multiplet $^2F_{5/2}$ of Ce$^{+3}$.
\end{abstract}


\maketitle

\section{Introduction}
Rare-earth intermetallic compounds containing Ce have attracted considerable attention due to their diverse properties including valence fluctuations, heavy-fermion behavior, and different types of magnetic ordering. The variety of different ground states arises due to the interaction of 4$f$ electrons with the crystal-electric field, and the competition between intersite Ruderman-Kittel-Kasuya-Yosida (RKKY) and onsite Kondo interaction.
The Ce-based Ce$M_x X_{4-x}$ compounds with tetragonal BaAl$_4$-type structure are model cases for this behavior as here, by changing the chemical composition or applying pressure, magnetic as well as nonmagnetic ground states can be obtained.\cite{Grin2,Flandorfer,Veronika} Their unit cell consists of a body-centered arrangement of Ce atoms in which the $M$ and $X$ ions occupy the remaining sites more or less randomly (see Fig.\,\ref{Fig1_Structure}). In the stoichiometric compounds with $x=1$, this atomic disorder can be removed by formation of a BaNiSn$_3$-type structure. This structure is a derivative of the BaAl$_4$ structure but has no inversion center due to the sequence of ordered Ce-X(1)-X(2)-M layers (see Fig.\,\ref{Fig1_Structure}). 

The recently discovered unconventional superconductivity in the non-centrosymmetric Ce$M$Si$_3$ and Ce$M$Ge$_3$ ($M$ = Co, Rh, Ir) has intensified the efforts to investigate the ground state of the related Ce$MX_3$ compounds with $M$ = Cu, Ni, Au, Pd, Pt and $X$ = Ga, Al.\cite{Kimura} 
CeAuGa$_{3}$, CeNiGa$_{3}$ and CePdGa$_{3}$ order antiferromagnetically at 2.4\,K, 1\,K, and 5.5\,K, respectively, \cite{Lohneysen,Cava,Kim} while CePtGa$_{3}$ shows a spin-glass type behavior.\cite{Steglich} All these compounds, in addition to their magnetic ordering, exhibit moderate heavy-fermion behavior. However, for several of them  conflicting reports about their magnetic behavior have appeared in the literature, in particular for samples with composition deviating from the stoichiometric $x=1$ composition. 

A model case for this aspect is the CeCuGa$_3$ alloy. The first report on CeCu$_x$Ga$_{4-x}$ compounds suggested ferromagnetic ordering of CeCuGa$_{3}$ at 3.5\,K.\cite{Grin} Studies on samples with varying Cu content, away from the stoichiometric composition, later indicated  that this order is stabilized by decreasing $x$.\cite{Sam1,Sam2,Oe} Mentink \textit{et al}.\cite{Mentink} and Sampathkumaran \textit{et al.},\cite{Sam2} on the other hand, suggested that CeCuGa$_{3}$ is paramagnetic down to 0.4\,K. A detailed investigation of polycrystalline CeCuGa$_{3}$ reveals Kondo-lattice behavior with magnetic ordering at 1.9\,K suggested to be antiferromagnetic.\cite{Martin} This was later supported by suceptibility measurements by Aoyama \textit{et al.}\cite{Aoyama,Kontani} reporting a higher transition temperature of 4\,K. From neutron-diffraction experiments on a single crystal (grown using the Czochralski method),\cite{Martin1} an incommensurate magnetic structure at 1.25\,K was inferred with a propagation vector $Q = (0.176, 0.176, 0)$. The magnetic scattering intensity extended up to 4\,K. In addition, a broad, somewhat ill-defined specific-heat anomaly was observed between 1.5\,K and 4\,K whereas in polycrystals with a reduced Cu content of $x=0.5$ a mean-field-like specific-heat anomaly was found.\cite{Sam2} 

In view of these conflicting reports on the type of magnetic ordering and in order to study the magnetic properties more precisely, we have grown single crystals of CeCuGa$_{3}$ from Ga flux and investigated the anisotropic magnetic properties with measurements of the AC and DC magnetization, specific heat, and electrical resistivity.

\section{Experimental Details}
Single crystals of CeCuGa$_{3}$ and LaCuGa$_{3}$ were grown by flux method using Ga as flux. The starting materials were high-purity La
and Ce (99.95\,\%), Cu (99.99\,\%) and Ga (99.999\,\%). Stoichiometric amounts of the constituents with excess of Ga (1:25) were put into
an alumina crucible and sealed in an evacuated quartz ampoule. The ampoule was heated to 1050\,$\mathrm{^{o}}$C over a period of
24\,hours and held at that temperature for another 24\,hours for proper homogenization. The furnace was then cooled down to 400\,$\mathrm{^{o}}$C at a rate of 1\,$\mathrm{^{o}}$C/h followed by fast cooling to room temperature. The crystals were separated from the flux by centrifuging. Energy-dispersive
X-ray analysis (EDAX) was performed on all crystals to identify their phase purity. The EDAX results showed 3\,\% to 6\,\% excess of Ga, which is attributed to the Ga flux used. X-ray powder diffraction patterns of all the compounds were recorded by powdering a small piece of single crystal (X-ray powder diffractometer STOE STADI P, Cu radiation). The single-crystal X-ray diffraction data sets were collected on an imaging-plate diffractometer system (STOE IPDS II Mo$K_{\alpha}$ radiation). The programs FULLPROF (Ref.~\onlinecite{FullProf}) and SHELXL (Shedrick, 1997) were used for a Rietveld analysis and structure refinement, respectively. The magnetic measurements were performed using a superconducting quantum interference device magnetometer from Quantum Design. The specific heat and resistivity were measured with a Physical Property Measurement System (PPMS) from Quantum Design.

\section{Experimental Results}
\subsection{Crystal structure}
\begin{figure}
\includegraphics[width=0.4\textwidth]{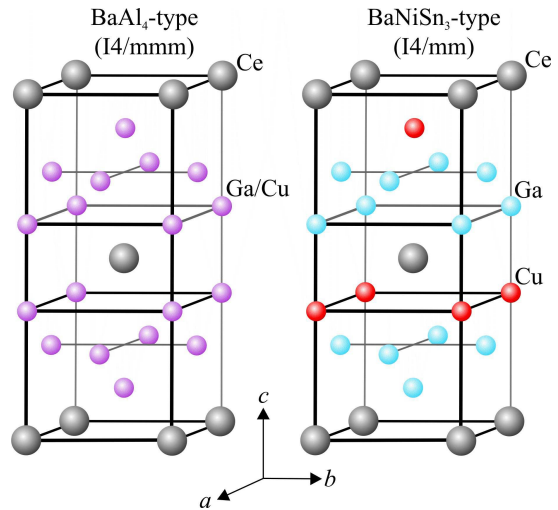}\caption{\label{Fig1_Structure}
(Color online) Two possible tetragonal crystal structures of $R$Cu$_x$Ga$_{4-x}$. If Cu and Ga occupy the 4$d$ and 4$e$ sites randomly the BaAl$_4$ structure is adopted. For the stoichiometric compound $x=1$ the Cu and Ga ions can order resulting in the related non-centrosymmetric BaNiSn$_3$-type structure.}
\end{figure}

$R$CuGa$_{3}$ compounds form in tetragonal derivatives of the BaAl$_{4}$-type structure (space group $I$/4$mmm$).\cite{Parthe} Depending on the degree of Cu-Ga order, these are mainly the BaAl$_{4}$, disordered ThCr$_2$Si$_2$ ($I$/4$mmm$), or non-centrosymmetric BaNiSn$_{3}$ structure ($I$/4$mm$) (see Fig.\,\ref{Fig1_Structure}). Powder diffraction is not able to distinguish between these structure types but provides the tetragonal lattice parameters which are for our LaCuGa$_{3}$ samples $a=4.32\,$\AA{} and $c = 10.436\,$\AA{} and for CeCuGa$_{3}$ $a = 4.273\,$\AA{} and $c=10.44\,$\AA. Together with the EDAX results, the lattice parameters demonstrate that the grown CeCu$_x$Ga$_{4-x}$ single crystals are, indeed, very close to the stoichiometric compound with $x = 1$.\cite{Grin,Sam1} The lattice parameter $a$ decreases and $c$ increases as we move from La to Ce. Overall, the unit-cell volume shrinks as expected from the lanthanide contraction. The $c/a$ ratio is $\approx$\,2.4 for both compounds indicating significant structural anisotropy. This is also reflected by the distance between nearest-neighbor Ce atoms which within the plane with 4.276\,\AA{} are much shorter than the  distance of 6.03\,\AA between Cu atoms of adjacent planes.

Single crystal x-ray diffraction has been used to resolve the structure type. In particular, we looked for a breakdown of the inversion symmetry by using the Flack parameter.\cite{Flack} As this parameter is not significantly enhanced we conclude that our single crystals adopt a BaAl$_{4}$ (or disordered ThCr$_2$Si$_2$) structure. In addition, small satellite Bragg peaks (with an intensity of 10$^{-3}$ of the main peaks) have been observed that point to a slight modulation of the structure. A similar observation was made by Martin \textit{et al.}\cite{Martin1} and attributed to a tendency towards an ordering of the Cu-Ga ions or a nearby structural transition.    

\begin{figure}
\includegraphics[width=0.5\textwidth]{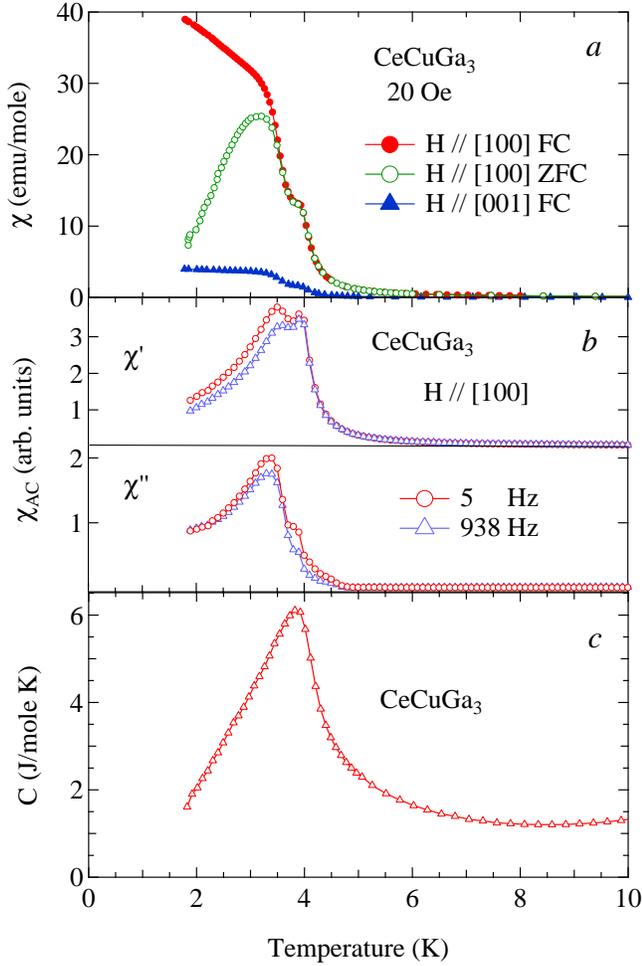} \caption{\label{Fig2_MT_Ce}
(Color online) (a) Magnetic susceptibility of CeCuGa$_{3}$ under zero-field-cooled (ZFC) and field-cooled (FC) conditions for $H \parallel \left[ 100 \right]$ and under FC condition for $H \parallel \left[ 001 \right]$ as a function of temperature. (b) AC susceptibility with $H \parallel \left[ 100 \right]$ and at frequencies of 5\,Hz and 938\,Hz. (c) Low-temperature specific heat of CeCuGa$_{3}$ showing the anomaly at the magnetic phase transition.}
\end{figure}

\subsection{Ferromagnetic ordering below 4~K}

Figure~\ref{Fig2_MT_Ce} shows the low-temperature DC and AC susceptibilities, $\chi_{DC}=M/H$ measured in a magnetic field $H = 20$\,Oe and $\chi_{AC}$, and the specific heat $C$ of CeCuGa$_3$. The data consistently indicate that CeCuGa$_3$ orders ferromagnetically at $T_{\rm{C}} \approx 4.2\,$K with the $ab$ plane as the easy plane of magnetization, cf. Fig.\,\ref{Fig2_MT_Ce}(a). The zero-field-cooled (ZFC) and field-cooled (FC) measurements of $\chi_{DC}$ for $H \parallel \left[ 100 \right]$, displayed in Fig.\,\ref{Fig2_MT_Ce}(a), demonstrate that with decreasing $T$, $\chi_{DC}$ reveals a Curie-Weiss-like increase followed by a kink at $T_{\rm{C}}$. 
At $T<T_{\rm{C}}$, the FC $\chi_{DC}$ continues to rise down to the lowest measured temperature of 1.8\,K. The splitting of the ZFC and FC curves below $T_C$ is in agreement with the expected behavior for a ferromagnet with domain-wall pinning which is enforced by the high magnetocrystalline anisotropy, i.e.,  $\chi_{DC}^{\left[ 100 \right]}/\chi_{DC}^{\left[ 001 \right]} \approx 8$ at $T=2\,$K, cf. Fig.\,\ref{Fig2_MT_Ce}(a).  An unusual feature of the $\chi_{DC}$ measurements is the appearance of a shoulder at temperatures above the bifurcation of the ZFC and FC curves for both field directions. This feature was observed for several samples prepared independently.

\begin{figure}
\includegraphics[width=0.5\textwidth]{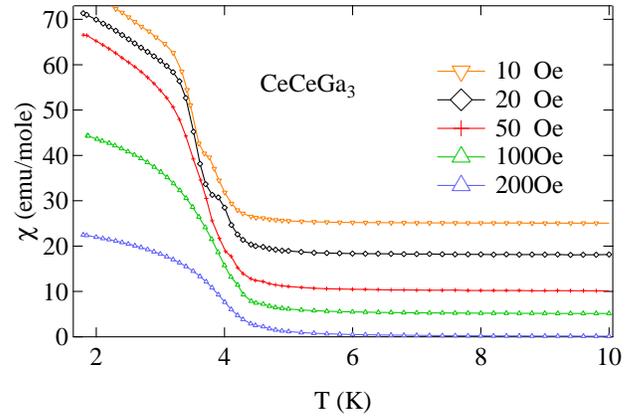}\caption{\label{Fig3_MT_H} (Color online) The susceptibility $\chi = M/H$ of CeCuGa$_{3}$ for $H \parallel \left[ 100 \right]$ as a function of $T$ in low magnetic fields. The curves are shifted with respect to another for a better readability.}
\end{figure}

\begin{figure}
\includegraphics[width=0.5\textwidth]{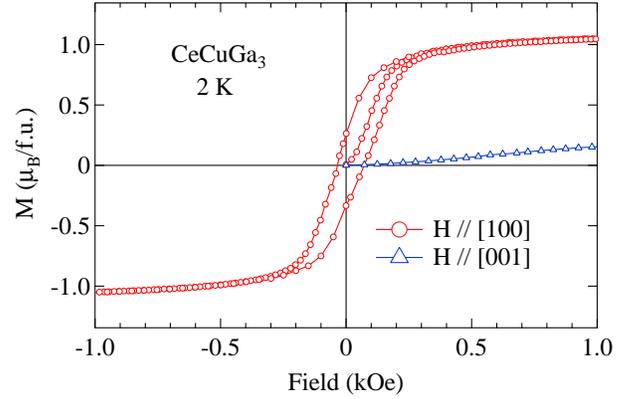}\caption{\label{Fig4_M_H} (Color online) Magnetic hysteresis loop $M(H)$ of CeCuGa$_3$ measured at 2\,K for $H \mid \mid [100]$. The low-field $(M)$ for $H \mid \mid [001]$ is also shown.}
\end{figure}

\begin{figure}
\includegraphics[width=0.5\textwidth]{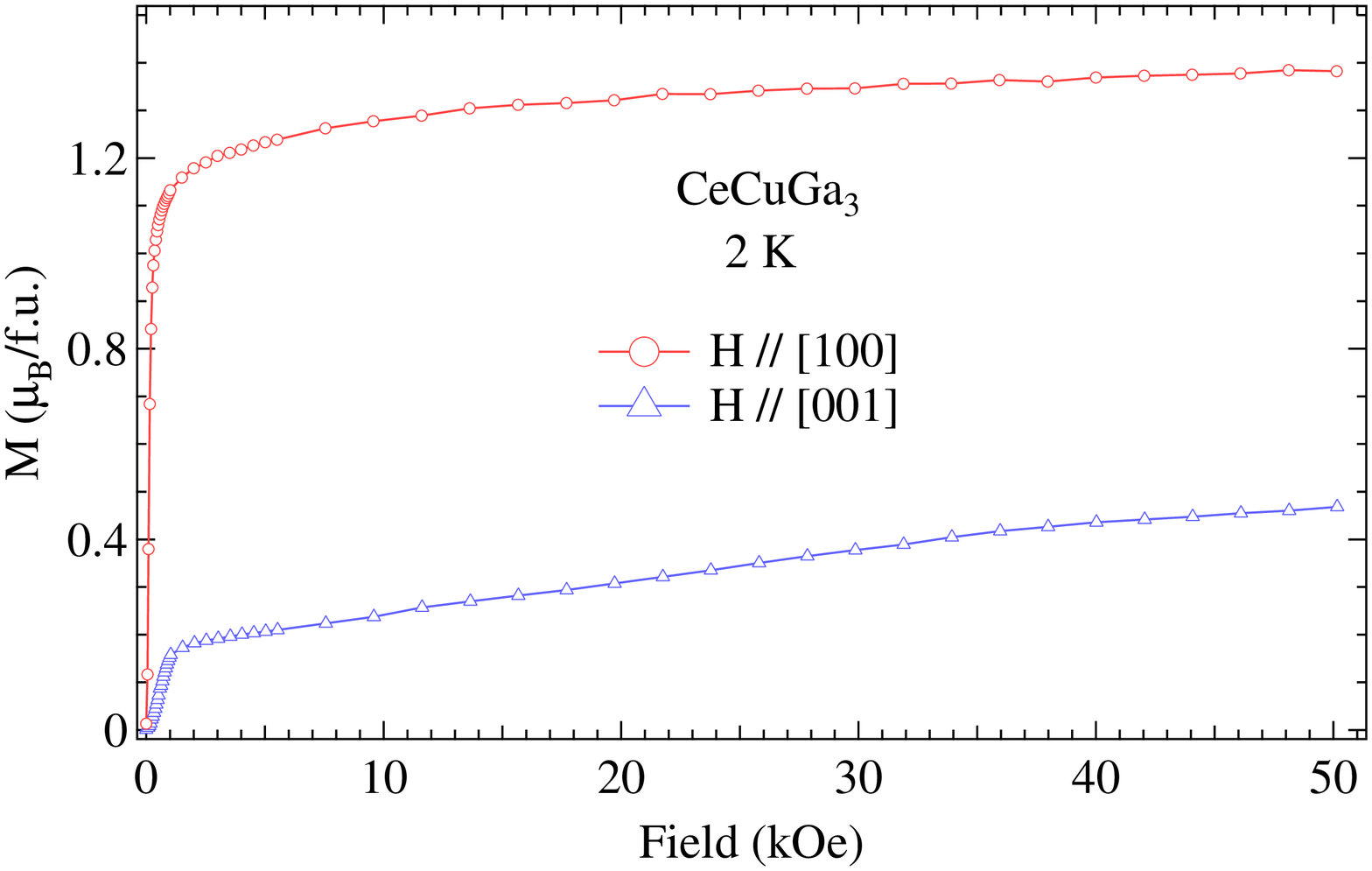}
\caption{\label{Fig5_MH} (Color online) Magnetic isotherms of CeCuGa$_{3}$ at 2\,K with field applied along the $\left[ 100 \right]$ and $\left[ 001 \right]$ directions.}
\end{figure}

To further investigate this anomaly, the AC susceptibility was measured at two frequencies of 5\,Hz and 938\,Hz with $H \parallel \left[ 100 \right]$ [see Fig.\,\ref{Fig2_MT_Ce}(b)]. In the real part of the AC susceptibility $\chi'_{AC}$, two clear peaks are visible at $\approx 3.5$\,K and $\approx 4$\,K. The imaginary part $\chi''_{AC}$ exhibits a peak and shoulder at the corresponding temperatures, indicating enhanced energy losses typical for thermodynamic phase or glass transitions. At the higher frequency of 938\,Hz, the high-temperature peak in $\chi'_{AC}$ remains unaltered but the intensity of the low-temperature peak is reduced and its position seems to be shifted towards higher temperatures compared to the 5-Hz data. This frequency dependence might suggest the onset of spin-glass behavior out of the ordered state towards lower temperatures. However, as the peaks in $\chi''_{AC}$ do not show such a dependence the shift in $\chi'_{AC}$ is most probably an artifact caused by the rising edge of the larger high-temperature peak in $\chi'_{AC}$.   

The specific heat of CeCuGa$_{3}$ is shown in Fig.\,\ref{Fig2_MT_Ce}(c) for comparison with the magnetic measurements. $C(T)$ shows a well-defined, rather sharp anomaly at the onset of the high-temperature shoulder in $\chi(T)$, which we therefore have taken as  $T_{\rm{C}}$. The size of the anomaly confirms bulk magnetic ordering. The specific heat does not show any feature corresponding to the low-temperature peak of $\chi_{\rm AC}(T)$ and the ZFC $\chi(T)$ for $H \mid \mid [100]$.

A systematic investigation of the field dependence of the $\chi(T) = M(T)/H$ at the double-step transition (Fig.\,\ref{Fig3_MT_H}) shows that the low-temperature shoulder shifts with field and eventually disappears at $H \approx 100\,$Oe. A hysteresis loop $M(H)$ measured at 2\,K for $H \mid \mid [100]$ do not show any signs of the anomaly (Fig.\,\ref{Fig4_M_H}).
Magnetic isotherms of CeCuGa$_3$ also measured at 2\,K with field applied along the $\left[ 100 \right]$ and $\left[ 001 \right]$ directions are displayed in Fig.\,\ref{Fig5_MH} and reflect the strong magnetic anisotropy.  From the high-field data at $H=50\,$kOe along the easy direction, we estimate a saturation moment of $M_s\approx 1.4\,\mu_{\rm{B}}$/Ce. The magnetization with field along the hard axis, $H \parallel \left[ 001 \right]$, undergoes a small jump at low fields and then increases slowly with $H$.

\begin{figure}
\includegraphics[width=0.5\textwidth]{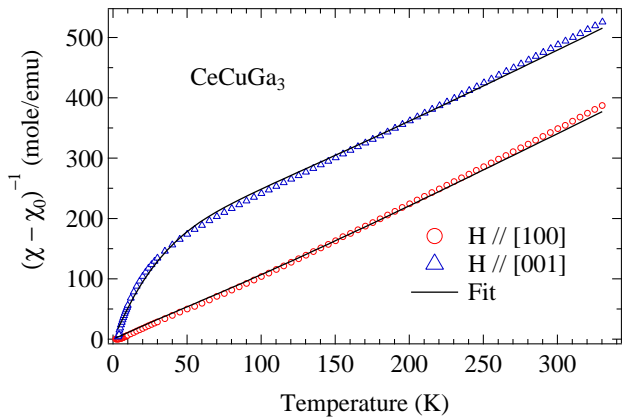} \caption{\label{Fig6_Inv_Chi}
(Color online) The inverse DC susceptibility of CeCuGa$_{3}$ after subtraction of  the temperature-independent electronic contribution, $(\chi-\chi_{0})^{-1}$, as a function of $T$ for $H \parallel \left[ 100 \right]$ and $\left[ 001 \right]$. The black lines are fits to the crystal-electric-field (CEF) model described in text.}
\end{figure}

\subsection{Crystal-electric-field effects}
\begin{figure}
\includegraphics[width=0.5\textwidth]{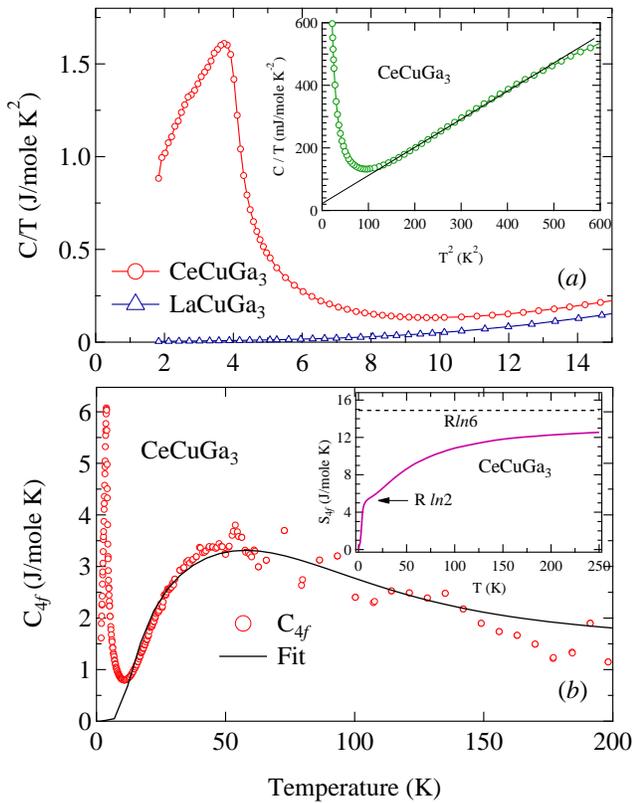}\caption{\label{Fig7_C} (Color online) (a) The specific heat of CeCuGa$_{3}$ and LaCuGa$_{3}$ plotted as $C/T$ vs. $T$. The inset shows the $C/T$ as a function of $T^{2}$. The linear fit was used to extract the Sommerfeld constant. (b) The $4f$ contribution to the specific heat of CeCuGa$_3$ with a fit to the Schottky anomaly as described in text. The inset shows the calculated $4f$ contribution to entropy.}
\end{figure}

The inverse magnetic susceptibility $\chi_{DC}^{-1}$ of CeCuGa$_{3}$ in the paramagnetic state is shown in Fig.\,\ref{Fig7_C}. A fit of a modified Curie-Weiss law for $T>70\,$K to the data for $H \parallel \left[ 100 \right]$ and $\left[ 001 \right]$ (not shown) yields effective moment $\mu_{eff}$, Curie temperature $\theta_{P}$, and temperature independent susceptibilities $\chi_{0}$ of 2.5\,$\mu_{\rm{B}}$/Ce and 2.51\,$\mu_{\rm{B}}$/Ce, 19\,K and -93\,K and -10$^{-4}\,$emu/mol and -7$\times 10^{-5}\,$emu/mol, respectively. The effective moments are close to the theoretically expected value for free Ce$^{3+}$ ions (2.54\,$\mu_{\rm{B}}$/Ce). The polycrystalline average of $\theta_{P}\approx -18.3\,$K seems  at first sight to be inconsistent with a ferromagnetic ground state. It has, however, taken into account that at lower temperatures the crystal electric field (CEF) acting on  the Ce ions lifts the degeneracy of the Hund's rule 4$f$ ground-state multiplet $^2F_{5/2}$, leading to distinct deviations from the Curie-Weiss behavior below 70\,K. A CEF calculation was done to obtain fits to the $\chi_{DC}^{-1}$ data for the whole temperature range 4 - 330\,K. The site symmetry of the Ce atoms in the CeCuGa$_{3}$ unit cell is supposed to be tetragonal (point symmetry $C_{4v}$). The corresponding CEF Hamiltonian is given by

\begin{equation}
\mathcal{H}_{{\rm CEF}}=B_{2}^{0}O_{2}^{0}+B_{4}^{0}O_{4}^{0}+B_{4}^{4}O_{4}^{4}\label{Eqn.1}
\end{equation}

\noindent
where $B_{\ell}^{m}$ and $O_{\ell}^{m}$ are the CEF parameters and the Stevens operators, respectively.\cite{Stevens,Hutchings} The CEF susceptibility is given by 

\begin{widetext}
\begin{equation}
\chi_{{\rm CEF}i}=N(g_{J}\mu_{{\rm B}})^{2}\frac{1}{Z}\left(\sum_{m\neq n}\mid\langle m\mid J_{i}\mid n\rangle\mid^{2}\frac{1-e^{-\beta\Delta_{m,n}}}{\Delta_{m,n}}e^{-\beta E_{n}}+\sum_{n}\mid\langle n\mid J_{i}\mid n\rangle\mid^{2}\beta e^{-\beta E_{n}}\right),\label{Eqn.2}
\end{equation}
 \end{widetext} 
\noindent
where $g_{J}$ is the Land\'e $g$\,-\,factor, $E_{n}$ and $\mid\! n\rangle$ are the $n$th eigenvalue and eigenfunction, respectively. $J_{i}$ ($i$\,=\,$x$, $y$ and $z$) are The components of the angular momentum, and $\Delta_{m,n}\,=\, E_{n}\,-\, E_{m}$, $Z\,=\,\sum_{n}e^{-\beta E_{n}}$ and $\beta\,=\,1/k_{{\rm B}}T$. The magnetic susceptibility including the molecular field constant $\lambda_{i}$ is given by
\begin{equation}
\chi_{i}^{-1}=\chi_{{\rm CEF}i}^{-1}-\lambda_{i}.\label{Eqn.3}
\end{equation}
The CEF fits to the inverse susceptibility data $(\chi - \chi_{0})^{-1}$ vs. $T$ are shown in Fig.\,\ref{Fig6_Inv_Chi}. The resulting CEF parameters are  $B_{2}^{0} = 11.0\,$K, $B_{4}^{0} = 0.127\,$K and $B_{4}^{4} = -3.0\,$K with molecular field constant $\lambda^{[100]} = 1.46\,$mol/emu and $\lambda^{[001]} = -1.9\,$mol/emu, respectively, for $H \parallel \left[ 100 \right]$ and $\left[ 001 \right]$. The positive value of the dominating CEF parameter $B_{2}^{0}$ is consistent with the $ab$ plane as the easy plane of magnetization. The CEF-split ground state multiplet $^2F_{5/2}$ of Ce$^{3+}$ ions in CeCuGa$_3$ can thus be described by three doublets with excitation energies of $\Delta_{1} = 50$\,K and $\Delta_{2} = 228\,$K.

Figure~\,\ref{Fig7_C}(a) shows the specific heat of CeCuGa$_3$ and its nonmagnetic analog LaCuGa$_3$ plotted as $C/T$ vs. $T$. The electronic contribution to the specific heat is estimated by extrapolating the "high-temperature" $T^2$ dependence of $C/T$ observed between 12 and 25\,K to $T = 0$ results into a Sommerfeld coefficient $\gamma \approx 20\,$mJ/mol K$^{2}$(inset of Fig.\,\ref{Fig7_C}(a)). This value is much smaller than that of 150\,mJ/mol K$^{2}$ reported by Martin \textit{et al.}\cite{Martin} for the antiferromagnetic CeCuGa$_3$. The magnetic contribution $C_{4f}$ to the specific heat (Fig.\,\ref{Fig7_C}(b) was determined from the $C(T)$ measurements by subtracting the phonon contribution estimated from the $C(T)$ of LaCuGa$_{3}$ and the electronic contribution $\gamma (T)$. The CEF contribution to $C$ with the CEF parameters extracted from the susceptibility measurements is shown in Fig.\,\ref{Fig7_C}(b). The good agreement with the measured $C_{4f}$ data confirms our calculated CEF level scheme and indicates that the broad high-temperature peak represents a Schottky anomaly due to CEF excitations. 

The calculated magnetic entropy $S_{4f}$ is shown in the inset of Fig.\,\ref{Fig7_C}(b). The magnetic entropy of the doublet ground state $R\ln2$ is reached slightly above $T_{\rm{C}}$. This points to a local-moment ferromagnet with only weak Kondo interaction as corroborated by the large ordered magnetic moment and the clear plateau of $S_{4f}(T)$ at $xR{\rm ln}2$ just above $T_c$. The total magnetic entropy at 250\,K is close to $R\ln6$. 

\begin{figure}
\includegraphics[width=0.5\textwidth]{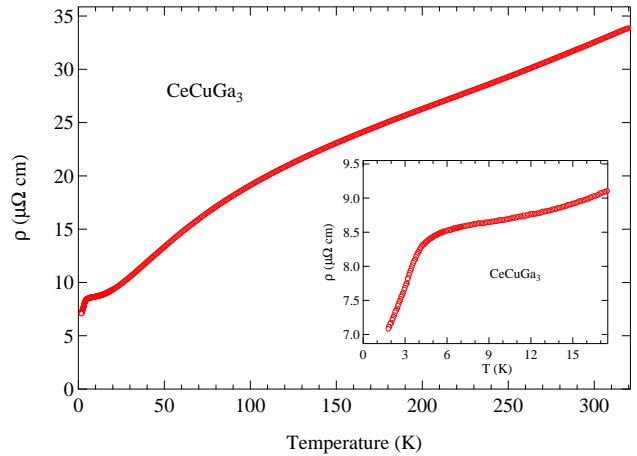}
\caption{\label{Fig8_rho} (Color online) Resistivity of CeCuGa$_{3}$ $\rho(T)$ and its counterpart LaCuGa$_{3}$ without 4$f$ electrons $\rho_{La}(T)$ which was used to estimate the magnetic contribution $\rho_{mag}\approx \rho - \rho_{La}$. The inset shows the low temperature resistivity of CeCuGa$_{3}$.}
\end{figure}

Finally the resistivity $\rho(T)$ of CeCuGa$_{3}$ is displayed in Fig.\,\ref{Fig8_rho}. With decreasing $T$, $\rho(T)$ exhibits a monotonic decrease followed by a leveling off around 10\,K and a drop at the ordering temperature of 4\,K (see inset of Fig.\,\ref{Fig8_rho}). These two latter features are attributed to a maximum of scattering by magnetic fluctuations at $T_C$. Hence no clear indication of a Kondo effect can be inferred from the $\rho(T)$ data. 
The non-linear decrease of $\rho(T)$ arises from a broad hump in $\rho(T)$ between 50\,K and 100\,K which can be attributed to scattering from crystal-field excitations.\cite{Cornut} This assignment is supported by a comparison to LaCuGa$_3$ which displays an almost linear $T$ dependence of $\rho(T)$ (not shown).

\section{Discussion}
Our measurements demonstrate that CeCuGa$_3$ has a ferromagnetic ground state if it adopts the centrosymmetric BaAl$_4$ structure. Its transition temperature matches the range of previously published measurements of CeCu$_x$Ga$_{4-x}$. According to the published structure investigations, all ferromagnetic alloys crystallize in the centrosymmetric BaAl$_4$-type structure. 
The examples for antiferromagnetic or incommensurate magnetic order reported so far have a Cu concentration of $x\approx 1$ and adopt the non-centrosymmetric BaNiSn$_3$-type structure. At higher Cu contents, for $1<x<1.5$, no long-range magnetic order could be observed down to 0.4\,K. 

As for $x=1$ different structural modifications can coexist, depending on the degree of atomic order it is conceivable that here multiple magnetic transitions occur and, due to the frustration caused by the competing interactions, glass-like behavior might appear. In this context, the double-step transition in our magnetization measurements and the superstructure peaks seen in the x-ray diffraction studies can be interpreted as first signs towards a partially ordered BaNiSn$_3$ structure which enhances the antiferromagnetic correlations. This additional contribution is, however, very small because in the specific heat no additional transitions could be identified. In contrast, Martin \textit{et al.}\cite{Martin1} reported more pronounced superstructure peaks and an unusual very broad anomaly in the specific heat that might perhaps be assigned to two different transitions at $\approx 4\,$K and $\approx 2\,$K, respectively. At the lower temperature of $\simeq$ 2\,K, they found the onset to the aforementioned, incommensurable, long-range magnetic order. 

The magnetic entropy as well as the Sommerfeld coefficient of CeCuGa$_3$ with BaNiSn$_3$, or BaAl$_4$ structure differ considerably from each other. While our measurements point to weak Kondo interactions (if present at all), Martin \textit{et al.}\cite{Martin} and Aoyama \textit{et al.}\cite{Aoyama} found a moderately heavy-fermion behavior with a clear Kondo-like minimum in the resistivity $\rho(T)$. Since the Kondo effect arises from  a local interaction between the 4$f$ electrons, and the conduction-band electrons this difference has to originate from differences in the immediate environment of the Ce$^{3+}$ ions.  
If in CeCu$_x$Ga$_{4-x}$ the Cu concentration is further increased long-range magnetic order vanishes. Sampathkumaran \textit{et al.}\cite{Sam1} observed a concomitant enhancement of the Kondo temperature. The border to the magnetic order in CeCu$_x$Ga$_{4-x}$ and most of the other aforementioned CeM$_x$X$_{4-x}$ alloys is close to $x\approx 1$, with the prospect to search for quantum phase transitions.

Our estimated CEF parameters are similar to those published by Oe \textit{et al.}\cite{Oe} for ferromagnetic CeCu$_{0.8}$Ga$_{3.2}$. The magnetic behavior, in particular the anisotropy and the saturation moment, can be well explained by the CEF splitting. To our knowledge there exists no detailed investigation of the CEF level scheme of CeCuGa$_3$ with BaNiSn$_3$ structure.

\section{Conclusion}
The magnetic behavior of single crystalline CeCuGa$_{3}$ has been studied. Depending on the Cu-Ga disorder the compound can adopt a tetragonal structure with (BaAl$_4$) or without inversion symmetry (BaNiSn$_3$). The magnetic properties of CeCuGa$_{3}$ and the isostructural Ce$M_x X_{4-x}$ alloys ($M$ = Ag, Au, Cu, Ni, Pd, Pt and $X$ = Al, Ga) sensitively depend on the environment of their Ce$^{3+}$ ions. If the composition is shifted away from the stoichiometric limit $x=1$, site disorder cannot be avoided and ferromagnetic ground states appear. In contrast to the non-centrosymmetric CeCuGa$_{3}$ without Cu/Ga site disorder, in the ferromagnetic compound only weak Kondo interactions could be observed. The magnetic anisotropy and the saturation moment correspond to the crystal-electric-field splitting of the Hund's rule 4$f$ ground-state multiplet. Although CeCuGa$_3$ seems to be close to the onset of magnetic order, so far no clear signs for quantum critical behavior or unconventional superconductivity could be found. Future investigations have to verify whether any of these features exist under hydrostatic pressure, as in the related, isostructural Ce$M$Si$_3$ and Ce$M$Ge$_3$ compounds ($M$ = Co, Rh, Ir).\cite{Kimura} 

\section{Acknowledgment}
D. A. Joshi acknowledges the post doc fellowship of the Karlsruhe Institute of Technology. Part of this work was supported by the Deutsche Forschungsgemeinschaft through the Research Unit FOR 960.

\end{document}